\documentclass[twocolumn]{aastex62}

\received{}
\revised{\today}
\accepted{}

\submitjournal{ApJL}

\shorttitle{High-Drag ISOs And Dynamical Streams}
\shortauthors{Eubanks}
\usepackage{gensymb} 
\usepackage{multirow}
\usepackage{amsmath}

\begin{document}

\title{High-Drag Interstellar Objects And Galactic Dynamical Streams}

\correspondingauthor{T. Marshall Eubanks}
\email{tme@asteroidinitiatives.com}
\author[0000-0001-9543-0414]{T.M. Eubanks}
\affil{Space Initiatives Inc, 
\\Clifton, Virginia 20124}

%% Note that the \and command from previous versions of AASTeX is now
%% depreciated in this version as it is no longer necessary. AASTeX 
%% automatically takes care of all commas and "and"s between authors names.

%% AASTeX 6.1 has the new \collaboration and \nocollaboration commands to
%% provide the collaboration status of a group of authors. These commands 
%% can be used either before or after the list of corresponding authors. The
%% argument for \collaboration is the collaboration identifier. Authors are
%% encouraged to surround collaboration identifiers with ()s. The 
%% \nocollaboration command takes no argument and exists to indicate that
%% the nearby authors are not part of surrounding collaborations.

%% Mark off your abstract in the ``abstract'' environment.

\begin{abstract}
%% Text of abstract
The nature of 1I/'Oumuamua (henceforth, 1I), the first interstellar object
known to pass through the solar system, remains mysterious. 
Feng \& Jones noted that the  incoming 1I  velocity vector ``at infinity'' (\textbf{v}$_{\infty}$) 
is close to the motion of the Pleiades dynamical stream (or Local Association), and suggested that 
1I is a young object ejected from a star in that stream. 
Micheli \textit{et al.} subsequently detected non-gravitational acceleration in the 1I trajectory; this acceleration would not be unusual in an active comet, but 1I observations failed to reveal any signs of activity.   
Bialy \& Loeb hypothesized that the anomalous 1I acceleration was instead due to radiation pressure, which would require an extremely 
low mass-to-area ratio (or area density).
Here I show that a low area density can also explain the very close kinematic association of 1I and the Pleiades stream, as
it renders 1I subject to drag capture by interstellar gas clouds. 
This supports the radiation pressure hypothesis 
and suggests that there is a significant population of low area density 
ISOs in the Galaxy, leading, through gas drag, to enhanced 
ISO concentrations in the galactic dynamical streams. 
Any interstellar object entrained in a dynamical stream will have a predictable 
incoming \textbf{v}$_{\infty}$; targeted deep surveys using this information
should be able to find dynamical stream objects months to as much as a year before their perihelion, providing the lead time needed for fast-response missions for the future \textit{in situ} exploration of such objects.
\end{abstract}

%% Keywords should appear after the \end{abstract} command. 
%% See the online documentation for the full list of available subject
%% keywords and the rules for their use.


%% From the front matter, we move on to the body of the paper.
%% Sections are demarcated by \section and \subsection, respectively.
%% Observe the use of the LaTeX \label
%% command after the \subsection to give a symbolic KEY to the
%% subsection for cross-referencing in a \ref command.
%% You can use LaTeX's \ref and \label commands to keep track of
%% cross-references to sections, equations, tables, and figures.
%% That way, if you change the order of any elements, LaTeX will
%% automatically renumber them.

\section{Introduction} 
\label{Sec:intro} 

 1I/'Oumuamua  was discovered near opposition on October 19, 2017 by \textit{Pan-STARRS1} at a distance  from Earth of $\sim$0.23 Astronomical Units (AU) \citep{Bacci-et-al-2017-a}. It was rapidly recognized as being on a strongly hyperbolic orbit, and given a new designation (1I/2017 U1) and the name 'Oumuamua. 1I does not exhibit the broad visual and near-IR absorption bands present in the spectra of many  asteroids \citep{Fitzsimmons-et-al-2017-a}, and so its composition remains very poorly constrained. 
Comparisons with stellar catalogs reveal that 1I has not passed extremely close to any star within the last few million years, and its original source system remains unknown \citep{Portegies-Zwart-et-al-2017-a,Gaidos-et-al-2017-a,Bailer-Jones-et-al-2018-a}.

Stellar perturbations make it hard to predict the detailed galactic trajectories of 
asteroid-sized InterStellar Objects (ISOs) over intervals much longer than a few million years \citep{Zhang-2017-a}. However, that does not mean that ISO velocities will become randomized  about the Local Standard of Rest (LSR). The dynamical LSR,  defined as the circular orbit velocity at the Sun's location, would be
the mean motion near the Sun for an axisymmetric galaxy. 
The Milky Way's velocity fields however are non-uniform,
with a substantial fraction of the stars in the solar neighborhood being concentrated 
in unbound collections of
stars called in this paper dynamical streams (but also known as associations or moving groups) 
\citep[see, e.g., ][]{Famaey-et-al-2005-a,Kushniruk-et-al-2017-a,Gaia-et-al-2018-b}.

\section{Stellar Streams in the Galactic Disk} 
\label{Sec:Stellar-Streams}
 
The study of the galactic stellar streams began in 1846 with the discovery of a distribution of stars sharing the proper motions of the Pleiades open cluster \citep{Kushniruk-et-al-2017-a}. For a long time it was thought that this and other streams were simply due to cluster evaporation (the gradual loss of stars from an open cluster), implying that the stars in the Pleiades stream should be no older than the cluster itself ($\sim$80 - 120 million years). As more data became available 
it became apparent that this was not so, with, for example, over half of the stars in the Pleiades stream being substantially older than the age of the Pleiades open cluster, rendering the evaporation model untenable \citep{Chereul-et-al-1998-b,Famaey-et-al-2008-a,Bovy-Hogg-2009-a}.

\cite{Antoja-et-al-2012-a}, using  RAVE spectroscopic survey data, found that 14.2\% of the stars in the solar neighborhood (out to 300 pc)
are in one of the five major streams considered in this paper, the 
Pleiades, Hyades, Sirius, Coma Berenices and Hercules streams, with another 
3.1\% of the stars belonging to 14 smaller dynamical streams.
 \cite{Francis-Anderson-2012-a} used 2MASS data to conclude that almost all local stars are  part of unbound kinematic streams, with the Pleiades stream being located in the leading edge of the Orion arm, and the Hyades stream being part of the Centaurus arm. 
The dynamical streams are likely associated with resonances in the Galaxy;
\cite{Michtchenko-et-al-2018-a} used \textit{Gaia} DR2 data to conclude that the 
Pleiades, Hyades and Coma-Berenices streams were all  associated with spiral-arm corotation resonances, while
the Sirius stream and Hercules stream are controlled by  Lindblad resonances.

\begin{figure}[!ht]
\begin{center}
\includegraphics[scale=0.75]{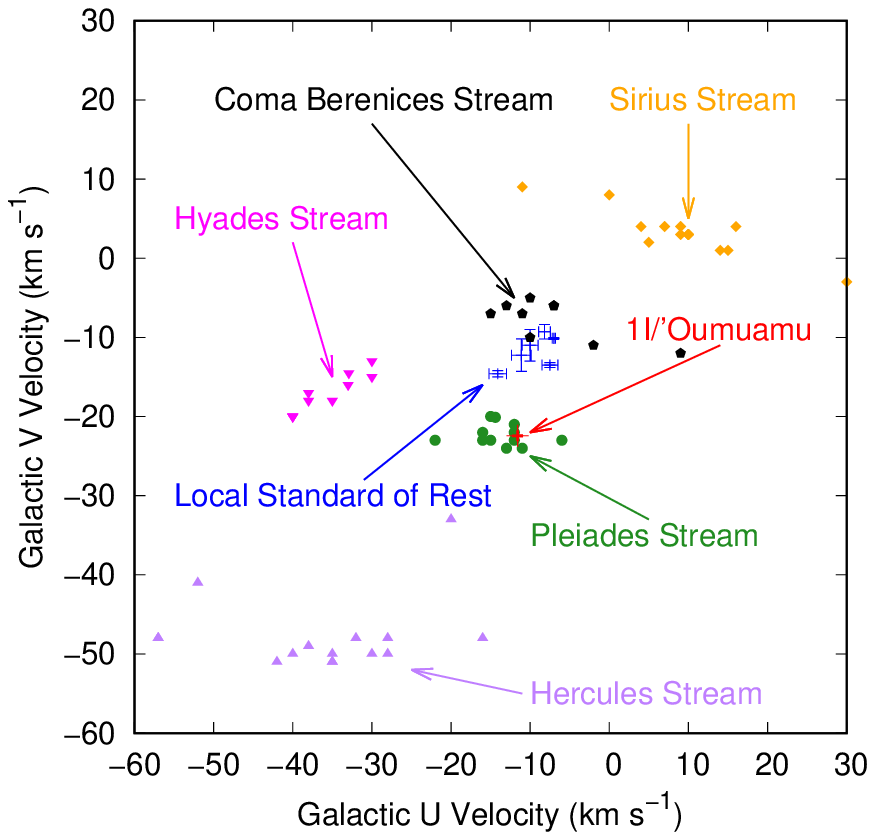}
\end{center}
\caption{The galactocentric U and V components of velocity for 1I, the LSR and the five largest local dynamical streams. The 1I incoming velocity is near the centroid of the determinations of the velocity of the Pleiades stream (Table \ref{table:galactocentric-velocities}). The stream velocity estimates are from the data and compilations in \citep{Kushniruk-et-al-2017-a}, supplemented by \citep{Chereul-et-al-1998-b,Liang-et-al-2017-a,Gaia-et-al-2018-b}. At least some of the scatter between the velocity estimates for individual streams seems to reflect substructure in the stream kinematics.
 The 1I inbound velocity is the average of the five 1I velocity solutions using anomalous acceleration models used by \citep{Bailer-Jones-et-al-2018-a}, with errors inflated to account for scatter in those solutions; the LSR velocity estimates are from \citep{Schonrich-et-al-2009-a,Francis-Anderson-2009-a,Francis-Anderson-2014-a,Huang-et-al-2015-a,Bland-Hawthorn-Gerhard-2016-a,Bobylev-Bajkova-2017-a}.
 }
\label{fig:UV-velocity}
\end{figure}

\section{1I and the Pleiades Stream} 
\label{Sec:1I-Pleiades}

\begin{figure}[!ht]
\begin{center}
\includegraphics[scale=0.75]{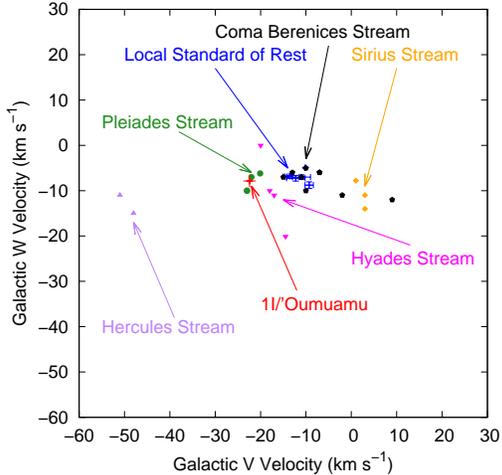}
s\end{center}
\caption{The data in Figure \ref{fig:UV-velocity}, but for the galactic V and W components of velocity. Not all stream surveys
report W, the component of velocity out of the galactic plane, and thus there are fewer stream data points in this image. 
The W component of these streams are all within 6 km s$^{-1}$ of the LSR, evidence that these are not tidal streams formed from galactic mergers, as those have much larger out of plane velocities \citep{Seabroke-et-al-2008-a}.
}
\label{fig:VW-velocity}
\end{figure}

Figures \ref{fig:UV-velocity} and \ref{fig:VW-velocity} show, for the galactocentric U-V and V-W planes respectively,  
velocity estimates for the LSR and the five major dynamical streams together with 
the mean 1I \textbf{v}$_{\infty}$ from \citep{Bailer-Jones-et-al-2018-a}.  
Table \ref{table:galactocentric-velocities} provides the mean galactocentric U,V,W components for all of these data;
all three components of the incoming 1I 
\textbf{v}$_{\infty}$ vector are close to the mean Pleiades stream velocity without any adjustment of biases.

\begin{deluxetable}{ccccc}
\tablecaption{Velocity Vectors in the Heliocentric Galactic Coordinate System
\label{table:galactocentric-velocities}}
\tablecolumns{5}
\tablehead{
\colhead{Velocity} & \colhead{U} & \colhead{V} & \colhead{W} & \colhead{$|$v$|$} }
%%                       &  km s$^{-1}$ &  km s$^{-1}$ &  km s$^{-1}$   &  km s$^{-1}$  \\
\startdata
1I \textbf{v}$_{\infty}$  & -11.6 $\pm$ 0.1  & -22.4 $\pm$ 0.1 & -7.9 $\pm$ 0.1 & 26.4 \\
LSR & -9.7 $\pm$ 2.7 &-11.8 $\pm$ 2.0 & -7.0 $\pm$ 1.2 & 16.7   \\
%% Local Stellar SD about LSR & 32.6 & 22.4 & 16.5 &  \\
\\
Sirius  &   5.2 $\pm$ 11.8 &  3.3 $\pm$ 2.8 & -8.2 $\pm$ 6.0 & 10.3 \\
Coma Beren. &  -7.3 $\pm$ 7.2 &  -7.8 $\pm$ 2.5 & -10.0 $\pm$ 2.0 & 14.6 \\
Pleiades  & -13.7 $\pm$ 3.8 & -22.3 $\pm$ 1.4 & -8.3 $\pm$ 2.0 & 27.5 \\
%% Pleiades Stream Member SD & 5.3 & 4.7 & 5.9 &  \\
Hyades  & -35.7 $\pm$ 4.0 & -17.2 $\pm$ 2.5 & -10.3 $\pm$ 8.2 & 40.3 \\
Hercules & -36.3 $\pm$ 12.0 & -47.7 $\pm$ 4.8 & -13.0 $\pm$ 2.8 & 61.4  \\
\\
1I - Pleiades & 2.1 $\pm$ 3.8 & -0.1 $\pm$ 1.4 & 0.4 $\pm$ 2.0 & 2.2   \\
\enddata
\tablecomments{Velocity Vector Components in the galactic U, V, W system, where U is radial (towards the Galactic Center), V is along the direction of galactic rotation, and W is orthogonal to the galactic disk. The final column provides the vector  magnitude. The data sources are described in the caption for Figure \ref{fig:UV-velocity}. The formal errors for 1I are inflated by the scatter of the solutions used in \citep{Bailer-Jones-et-al-2018-a}, the other formal errors are the rms scatter of the data plotted in Figures \ref{fig:UV-velocity} and \ref{fig:VW-velocity}. 
}
\end{deluxetable}

\begin{figure}[!ht]
\begin{center}
\includegraphics[scale=0.75]{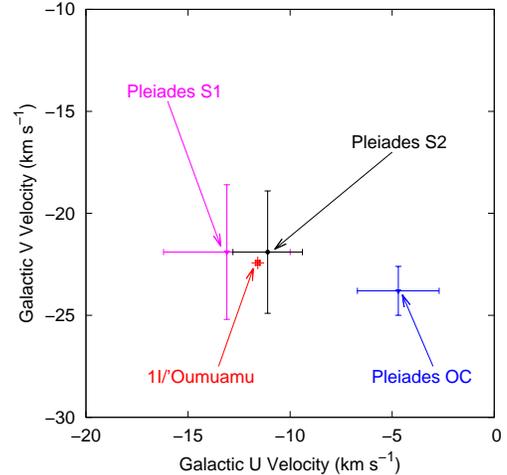}
\end{center}
\caption{1I data as in Figure \ref{fig:UV-velocity} compared with the fine scale divisions of the Pleiades stream derived by \citep{Chereul-et-al-1998-b,Chereul-et-al-1999-a} from Hipparcos data. The OCl represents a younger stream apparently populated with stars from the Pleiades open cluster, while S1 and S2 are 
substreams of the older SCl stream.  The error bars for the stream data are the stellar velocity dispersion of the indicated streams, while the error bars for the 1I data are based on the rms scatter of the 
various solutions in \citep{Bailer-Jones-et-al-2018-a}.  
 }
\label{fig:UV-stream-split}
\end{figure}C

\subsection{1I and the Pleiades Substreams}
\label{1I-and-Pleiades-Substreams}

A high resolution study of the Pleiades stream using Hipparcos data \citep{Chereul-et-al-1998-b,Chereul-et-al-1999-a} found two major components to that stream, which they labeled the Open Cluster (OCl) and Super Cluster (SCl) streams,
with the OCl stream being associated with the Pleiades star formation region. 
The SCl stream is divisible into two finer-grained substreams, S1 and S2, which are kinematically adjacent but contain stars of different origin and ages. Figure \ref{fig:UV-stream-split} shows the
U and V components of these velocities; the error bars on each substream velocity component being the rms 
of that velocity component for the 
stars in that substream.
The observed 1I \textbf{v}$_{\infty}$ velocity clearly favors its membership in the SCl over the OCl; 
the magnitude of the 3-dimensional separation of the 1I  \textbf{v}$_{\infty}$ 
and the stream velocity centroid is $<$  2 km s$^{-1}$ for both the S1 and S2 streams, substantially less than the $\sim$ 3 km s$^{-1}$ rms velocity dispersions of these streams. The chance that a randomly selected velocity would fall within 2 km s$^{-1}$ of one of 19 streams in the available  3-D velocity space is $<$ 10$^{-3}$, strong evidence that 1I was entrained in the SCl stream, but not proof that it originated in a star system in that stream.

\subsection{Gas Drag Capture of Low-$\beta$ ISOs}

Recent research indicates that 
1I had a small, but highly significant ($\sim$ 30 $\sigma$), anomalous acceleration during its period of observation (October 14th, 2017 - January 2nd, 2018), the observed non-gravitational acceleration being predominately radial and declining with distance from the Sun \citep{Micheli-et-al-2018-a}. Anomalous acceleration in small solar system objects can be caused by cometary activity, but no outgassing was detected from 1I, with in particular very low limits being set 
on its dust, CO and CO$_{2}$ emissions
by the \textit{Spitzer} Space Telescope \citep{Trilling-et-al-2018-a}. In addition, even a small assymmetry in the required thrust would strongly torque an elongated body the presumed size of 1I, causing faster than observed rotational variations \citep{Rafikov-2018-a}.  

\cite{Bialy-Loeb-2018-a} proposed instead that the 
1I anomalous acceleration  was due to Solar radiation pressure, which 
functionally fits the observed acceleration signature. 
This solution, however, requires 
a mass-to-area ratio, $\beta$, of 0.93 $\pm$ 0.03 kg m$^{-2}$, much lower than the $\beta$ for any known asteroid, and comparable to the area density of a light-sail, leading to speculation that 1I could be of artificial origin. \cite{Moro-Martin-2019-a} showed that similarly low area densities could also be obtained from a porous icy aggregate formed outsde the snowline of a protoplanetary disk

Although there is no consensus about the nature of 1I, and many
researchers prefer a cometary model with dust-free outgassing \citep{Micheli-et-al-2018-a,Sekanina-2019-a}, it is worth considering the observational consequences of a population of low-$\beta$ ISOs.
The trajectories of ISOs with $\beta$ $\sim$ 1 kg m$^{-2}$ will be significantly affected by drag in the InterStellar Medium (ISM). The Newtonian drag equation \citep{Moe-et-al-1995-a,Scherer-2000-a} is
\begin{equation}
\frac{d\mathrm{v}}{dt} \sim - \frac{1}{2}\ \frac{\mathrm{C}_{\mathrm{D}}\ \mathrm{v}^{2}\ \rho_{\mathrm{ISM}}}{\beta}    \ ,
\label{eq:drag-eq}
\end{equation}
where 
C$_{\mathrm{D}}$ is the dimensionless drag coefficient (typically $\sim$2.6 for Earth satellites), 
v the magnitude of the relative ISO-ISM velocity and
$\rho_{\mathrm{ISM}}$ the ISM density. If the ISM is assumed to be predominately atomic Hydrogen,
\begin{equation}
\rho_{\mathrm{ISM}}\ =\ \left(\frac{\mathrm{n}_{0}}{1\ \mathrm{cm}^{-3}}\right)\ 
\times 1.7\ \times\ 10^{-21}\ \mathrm{kg}\ \mathrm{m}^{-3} , 
\label{eq:rho-eq}
\end{equation}
n$_{0}$ being the particle density. In a region of space with a  constant $\rho_{\mathrm{ISM}}$, 
the distance, L$_{\mathrm{D}}$, for a factor of 2 reduction in velocity is 
\begin{equation}
    \begin{split}
    	 \mathrm{L}_{\mathrm{D}} & = \frac{2\ \beta}{\rho_{\mathrm{ISM}} \mathrm{C}_{\mathrm{D}}} \\
							 	 & \sim 4 \times 10^{4}\ \mathrm{lyr}\ \left( \frac{1\ \mathrm{cm}^{-3}}{\mathrm{n}_{0}}\right)
 \left( \frac{\beta}{0.93\ \mathrm{kg}\ \mathrm{m}^{-2}}\right) \\
    \end{split}
\label{eq:characteristic-length}
\end{equation}
In the background galactic disk, the ISM n$_{0}$ is typically $\lesssim$ 1 cm$^{-3}$ \citep{Scherer-2000-a}, so that 
objects with 1I-type $\beta$ should,  by Equations \ref{eq:drag-eq} and \ref{eq:characteristic-length},  be able to travel 
across much of the galactic disk without losing much of their peculiar velocity.

A small spherical asteroid or comet with a radius R and a uniform density $\rho$ would have
\begin{equation}
\beta\ \sim\ \left(\frac{\mathrm{R}}{100\ \mathrm{m}}\right)\ 
\left(\frac{\rho}{1000\ \mathrm{kg}\ \mathrm{m}^{-3}}\right)\ 
\times\ 10^{5}\ \mathrm{kg}\ \mathrm{m}^{-2}\ ,
\label{eq:low-drag}
\end{equation}
An object with a typical cometary or asteroid density of $\sim$500 - 3000 kg m$^{-3}$ \citep{Carry-2012-a} 
would need a radius 
$\lesssim$ 1 mm to have a $\beta$ comparable to 1I's. 
Solar system asteroids and comets are thus high-$\beta$ objects; ours and other planetary systems must have ejected large numbers of high-$\beta$ planetesimals, asteroids and comets over the 
course of their histories \citep{Engelhardt-et-al-2017-a}. 
If 1I truly is a low-$\beta$ object, there therefore must be two populations of ISOs in the 100-meter size range, one with area ratios similar to solar system asteroids and negligible drag even in the densest molecular clouds, and the other, possibly more numerous, being 1I-type objects sensitive to ISM drag.

The galactic spiral arms and their dynamical streams contain
stellar nurseries with relatively high gas densities \citep{Francis-Anderson-2012-a}.
Equation \ref{eq:characteristic-length} indicates that a star formation region such as the Orion Nebula (M42), with a central gas density $\sim$ 10$^{4}$ cm$^{-3}$ \citep{Johnson-1961-a}, 
should be able to capture a low-$\beta$ ISO through gas drag. 
The mean time between ISO-cloud interactions in the galactic disk is thought to vary from $\sim$30 Myr for HI regions to
$\sim$1 Gyr for Giant Molecular Clouds \citep{Yeghikyan-Fahr-2003-a}. The presence of 1I in the SCl instead of the OCl stream suggests that it may have been slowed by a gas cloud, possibly a stellar nursery, predating the Pleiades open cluster; finding and dating star formation regions in the SCl stream could thus potentially provide a lower bound for 1I's age.

\section{Efficient Searches for Galactic Stream Asteroids} 
\label{Sec:Finding-ISOs}

ISOs from a given dynamical stream will appear to enter the solar system from a specific radiant in the sky; Figure \ref{fig:Sky-plot} shows the radiants for the 5 major streams considered in this paper. As seen from the Earth, an incoming ISO will execute an expanding parallactic spiral centered around its radiant; preperihelion detection of incoming ISOs is thus possible using deep surveys centered about the stream radiants \citep{Eubanks-2019-a}.

\subsection{Number Density of Interstellar Asteroids} 
\label{SubSec:number-density}

\textit{Pan-STARRS1} detected 1I after only 3.5 years of observing in its current survey mode.  \cite{Do-et-al-2018-a} calculated that in that period
\textit{Pan-STARRS1} scanned $\sim$5 AU$^{3}$, implying \citep{Do-et-al-2018-a,Trilling-et-al-2017-a} 
an upper limit on n$_{\mathrm{IS}}$, the ISO number density, of
\begin{equation}
\mathrm{n}_{\mathrm{IS}}\ \lesssim\ 0.2\ \mathrm{AU}^{-3}\ .
\label{eq:number-density}
\end{equation}
If the upper bound in Equation \ref{eq:number-density} is indicative of the density of $\sim$100-meter sized ISOs in the galactic disk, then these objects must be very common, with  $\sim$10$^{16}$ such objects for each star in the Galaxy \citep{Engelhardt-et-al-2017-a,Raymond-et-al-2017-a,Do-et-al-2018-a}. 

A common means of extending number density estimates is through a power law model,  
where the cumulative density \citep{Engelhardt-et-al-2017-a} is
\begin{equation}
\mathrm{n}_{\mathrm{IS}}(\mathrm{Diameter}\ \geq\ \mathrm{D}) \propto \mathrm{D}^{-\alpha_{\mathrm{IS}}}  \ ,
\label{eq:N_IS_cum}
\end{equation}
%% \mathrm{IS}
 for a body of diameter D, $\alpha_{\mathrm{IS}}$ being the power law exponent. 
There are only very weak constraints on 1 km diameter ISOs, particularly if they are assumed to be inactive. 
\cite{Engelhardt-et-al-2017-a} obtained a firm observational upper bound for 
inactive ISOs of n$_{\mathrm{1 km}}$ $\lesssim$ 1.4 $\times$ 10$^{-2}$ AU$^{-3}$, corresponding to 
 $\alpha_{\mathrm{IS}}$ $\gtrsim$ 1, but there is no consensus about the lower limit of n$_{\mathrm{1 km}}$.
 Theoretical estimates of ISO production from before the discovery of 1I tend to strongly underpredict the rate of 100-meter sized ISOs, and thus may underpredict larger bodies as well, while constraints based on total mass 
 production do not apply because of the substantially lower mass of low-$\beta$ ISOs (and, of course, how $\beta$ might scale with mass is also unknown). 
 
The upper limit of Equation \ref{eq:number-density}  predicts that there should be several 1I-sized ISOs inside the Earth's orbit in the course of a year \citep{Meech-et-al-2017-a}, while the \citep{Engelhardt-et-al-2017-a} 
upper bound corresponds to roughly 1 km-sized ISO passing within the Earth's orbit per year. 
A decade-long deep survey should thus be able to either detect km-sized ISOs or 
limit $\alpha_{\mathrm{IS}}$ to be $\lesssim$ 2. If these objects are predominately entrained in the galactic dynamical streams, 
targeted deep optical surveys should be able to 
find them well before perihelion, providing a sufficiently long observation arc to enable the determination of
both the fraction of ISOs with cometary activity and the mass-area ratios of these objects. 

\subsection{Deep Surveys for Stream ISOs} 
\label{SubSec:Deep-Surveys}

\begin{figure}[!ht]
\begin{center}
\includegraphics[scale=0.7]{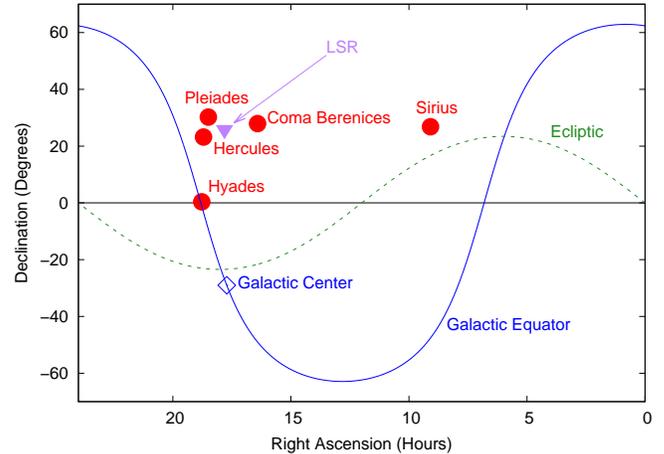}
\end{center}
\caption{Incoming radiants of the 5 largest galactic streams in the solar neighborhood using the galactocentric velocities in Table \ref{table:galactocentric-velocities}, together with the Solar Apex (the incoming LSR radiant).
The dispersion in the stream velocities as seen in the \textit{Gaia} DR2 is comparable to or smaller than the size of the symbols; a substantial fraction of the stars in the solar neighborhood belong to one of these streams, and it is thus reasonable to assume that a substantial fraction of incoming ISOs will come from these radiants. 
}
\label{fig:Sky-plot}
\end{figure}

Figure \ref{fig:Sky-plot} shows the mean radiants (the directions of their approach before solar perturbations) 
for each of the five major streams shown in 
Table \ref{table:galactocentric-velocities}. 
\cite{Seligman-Laughlin-2018-a} used an ellipsoidal distribution for stellar velocities and predicted that ISOs were more likely to come from a very broad angular region centered on the LSR, which the dynamical stream
model resolves into a much more angularly constrained set of stream radiants.

It should be possible to discover ISOs approaching from known directions using existing telescopes. 
The 8-meter Subaru telescope with its wide-field Hyper Suprime-Cam (HSC) camera \citep{Miyazaki-et-al-2018-a}  would have been able to detect 1I at M $\sim$ 26.75 at the beginning of June, 2017, 3 months before its perihelion and 4.5 months before its discovery, providing several months of lead time for the hypothetical fly-by mission described in \citep{Seligman-Laughlin-2018-a}. The same limiting magnitude would suffice to discover a ten times larger-diameter body with the same albedo (i.e., one with an absolute magnitude H $\sim$ 17)  a year before perihelion at a distance of $ \sim$8 AU. 

Deep searches for incoming ISOs are thus possible with existing telescopes. Roughly 75 square degrees would have to be covered to completely scan the Pleiades radiant for incoming ISOs one year in advance, with the total exposure time
to image each radiant to a magnitude M = 26.75 (based on the HSC Exposure Time Calculator) varying from 6 to 18 hours, depending on the phase of the Moon. The Pleiades and Hercules stream radiants are only 
separated by $\sim$7.5\degree; a survey of the Pleiades radiant would thus include a substantial fraction of the Hercules radiant, and could detect incoming ISOs from that stream as well.

\section{Discussion and Conclusions} 
\label{Sec:Conclusions}

The presence of 1I near the kinematic center of the Pleiades stream  
suggests that it is a low-$\beta$ object subject to ISM drag. 
The association of interstellar asteroids and galactic streams  hypothesized in this paper can be directly tested
through the search for, and discovery of, more ISOs. In particular, the Pleiades stream apparently only contains about 3.2\% of the stars in the solar neighborhood \citep{Antoja-et-al-2012-a}. Whether the discovery of the first ISO from the Pleiades stream was simply a matter of chance,
or whether that stream is for some reason especially rich in low-$\beta$ ISOs,
will be straightforward to determine with additional ISO discoveries. 

The discovery, observation and eventual exploration of ISOs passing through the solar system offers a profound opportunity to determine both the physical properties of these bodies and their role in the dynamics and evolution of the Galaxy.
The discovery of only a few additional ISOs will substantially 
reduce the uncertainties in their number density spectrum. 
Even with short observational arcs it should 
be possible to determine which objects have been entrained into a galactic stream, and 
thus possibly to distinguish between low and high $\beta$ objects even without the direct detection of anomalous accelerations. 
For ISOs discovered before perihelion,
it should be possible to detect or severely limit both activity and non-gravitational acceleration, and thus
determine whether any anomalous acceleration is due to outgassing or to radiation pressure. 

Directed ISO searches should increase their discovery rate. 
A targeted search of the radiants of the Pleiades and Hercules streams (which are close together in the sky) might be able to scan 20 AU$^{3}$ yr$^{-1}$ for 1 km-size bodies. If 1I represents a dense population of ISOs in the Pleiades stream, than such a survey might detect several ISOs per year. 
If, on the other hand, ISOs are distributed in the same proportion as local stars, then roughly 3.2\% and 1.4\% 
of the incoming ISOs would be from the Pleiades and Hercules streams, respectively; the same survey would yield on average one detection every 6 years. 
An on-going survey targeted on dynamical stream ISOs has a decent chance at detecting these objects well
before their perihelion passage, providing the lead time needed for fast-response missions for the \textit{in situ} exploration of these interstellar bodies.

Small objects in interstellar space are probes of the dynamics of the Galaxy. 
A low-$\beta$ ISO, once ejected from its source system,
should orbit the Galaxy until it reaches a turbulent region with high gas density, whereupon it would be likely to stop and join the bulk motion of the gas, and thereafter (if it was not trapped or destroyed in a new stellar system) be released as part of a dynamical stream. This appears to be 
1I's likely nomadic history. This hypothesis could be directly tested by sending a mission to 1I \citep{Hein-et-al-2017-a} or by sending a future mission to other ISOs as they pass through the solar system \citep{Seligman-Laughlin-2018-a}. 

If low-$\beta$ ISOs are indeed common, a population of these objects 
could have been captured by gas drag during the nebula stage of the formation of the solar system \citep{Grishin-et-al-2018-a} and 
retained in the outer solar system today (radiation pressure would prevent a low-$\beta$ object from having a stable orbit in the inner system). This population would be easily distinguishable from ISOs captured by three body gravitational interactions after the formation of the solar system 
\citep{Siraj-Loeb-2018-a}, and from low-$\beta$ objects temporarily captured close to the Sun by solar radiation pressure perturbations.
Primordially captured ISOs, perturbed into the inner solar system by the 
mechanisms that produce long-period comets, should thus be searched for 
as small inactive comets  on nearly-hyperbolic trajectories with unexpectedly large non-gravitational accelerations.

\acknowledgments
\section{Acknowledgments}
The author acknowledges suggestions from the anonymous referees.  
This work used the JPL 
HORIZONS system and Small-Body Database Browser, the NASA/IPAC Extragalactic Database (NED), the Minor Planet Center database and the SIMBAD data base operated at CDS, Strasbourg, France. The author was supported by Space Initiatives and the Institute for Interstellar Studies (I4IS).

%%\appendix

%%\section*{References}

%%\begin{thebibliography}{}
%%end{thebibliography}

%% References with bibTeX database:

\bibliographystyle{aasjournal}
%% \bibliography{../eubanks}

\end{document}